# Wearable Haptic Device to Render 360-degree Torque Feedback on the Wrist

Seungchae kim[1], Mohammad Shadman Hashem[1], and Seokhee Jeon[1]

[1] Department of Computer Science and Engineering, Kyung Hee University, Seoul, Republic of Korea

(Email: jeon@khu.ac.kr)

**Abstract ---** Haptic feedback increases the realism of virtual environments. This paper proposes a wearable haptic device that renders torque feedback to the user's wrist from any angle. The device comprises a control part and a handle part. The control part consists of three DC gear motors and a microcontroller, while the handle part securely holds the Oculus Quest 2 right controller. The control part manages string tension to deliver the sensation of torque feedback during interactions with virtual tools or objects. The three points of the handle part are connected to the three motors of the control part via strings, which pull the handle part to render precise 360-degree (yaw and pitch) torque feedback to the user's wrist. Finally, to show the effectiveness of the proposed device, two VR demos were implemented- Shooting Game and Shielding Experience.

## 1 INTRODUCTION

Virtual reality (VR) provides visual and auditory feedback, allowing people to explore virtual environments [1] However, without haptic feedback, VR environments may lack a sense of reality and immersion. Physical interaction signals increase the sense of reality and immersion in VR environments in terms of haptic feedback [2]. Haptic feedback implements physical interaction signals similar to various real-life experience such as pressure, vibration, stiffness, friction, heat, force, and torque in VR environments [3]. This paper focuses on the torque feedback of the user's wrist among these various haptic feedbacks.

Previous studies tend to focus mainly on implementing kinesthetic force feedback on the entire arm or finger [4]. However, kinesthetic torque feedback on the wrist is comparatively less explored [5]. Studies on haptic devices [6-8] capable of providing torque feedback to the wrist have limitations, such as being heavy and bulky, having delays in rendering torque feedback, or generating torque in only a limited direction.

In this paper, a simple, lightweight, and highly portable wearable device was proposed and designed that render yaw and pitch direction torque feedback on the wrist using three DC gear motors and three strings only. The proposed haptic device provides torque feedback along two primary axes, yaw (horizontal) and pitch (vertical), effectively covering 360 degrees of interaction as Figure 1.

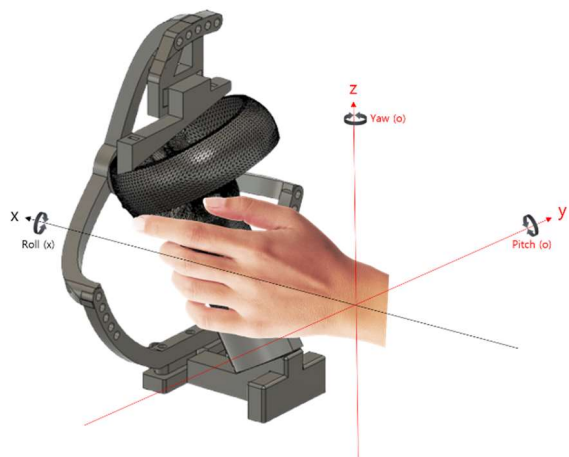

Fig.1  The proposed haptic device delivers torque feedback along the yaw (horizontal) and pitch (vertical) axes

The device consists of two parts. One is the handle part, and the other is the control part Figure 2. The handle part and the control part are connected through three nylon threads connected to three DC gear motors. The gear motor generates torque feedback by pulling the nylon thread.

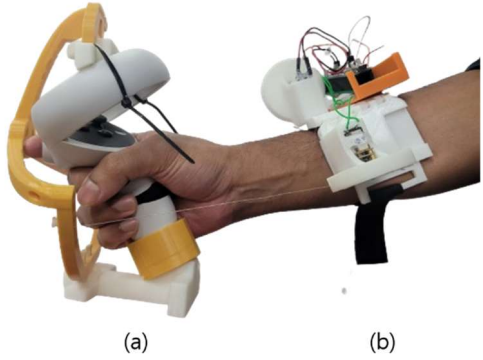

Fig.2  Overall design of the proposed haptic device; (a) Handle part, (b) Control part

## 2 PROPOSED SYSTEM

### 2.1 Virtual Application Scenario

Two VR demos were implemented to show the effectiveness of our proposed device. The demos were designed using the Unity Game Engine. The first demo is a shooting game that allows the users to perceive the sensation of gun shooting using three different kinds of guns: a pistol, a rifle, and a shotgun. The second demo is based on a shielding experience that allows the users to perceive the resistive sensation of shielding results from colliding bullets from different directions. Meta Quest2 VR device was used for the VR platform.

In the shooting demo, Figure 3, the torque sensation resulting from the gun recoiling action is mainly perceived in the direction of 90 degrees. Therefore, the center thread is mainly used to generate torque in the direction of 90 degrees on the user's wrist in the shooting demo. In the shooting demo, users can perceive the torque sensation that varies depending on the gun.

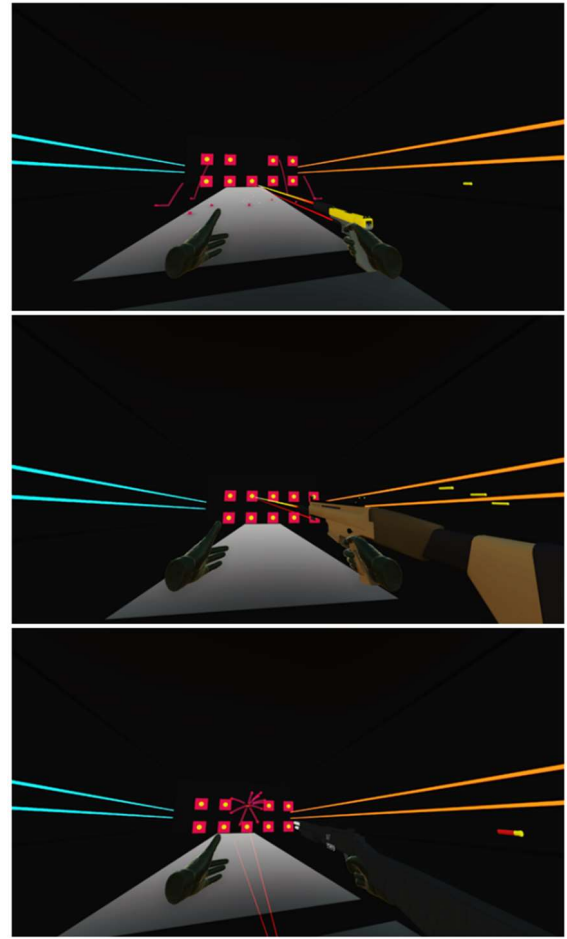

Fig.3  Shooting different three guns (a pistol, a rifle, and a shotgun) in the shooting demo.

In Shielding Demo, Figure 4, the user uses a shield to block incoming bullets from different directions. There are two types of bullets in the shielding demo: red and blue. The blue bullet is larger than the red bullet and has a greater destructive power during collision with the shield. When a user blocks a bullet using a translucent shield, a mark is briefly shown on the surface of the shield of that blocked bullet, allowing the user to know which part of the shield blocked the bullet. Depending on the position of the blocked bullet, users perceive the torque sensation on the wrist in different directions.

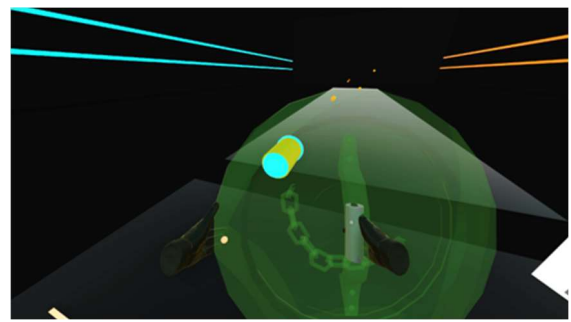

Fig.4  Blocking a bullet with a shield in the shielding demo.

## 3 Conclusion

In this paper, a portable haptic handle device was proposed and designed that renders 360-degree (yaw and pitch) torque feedback to the user's wrist. This device has a simple and lightweight structure, increasing the applicability in various VR environments. By controlling the three strings in the three given directions, torque feedback was implemented in all directions, and vibration feedback was also able to render through the existing device. The device improves the sense of immersion and reality of the VR experience.

In future research, the weight of the device will be further reduced and the design will be more compact. In addition, it will be improved to provide various types of torque feedback suitable for different scenarios such as playing tennis, swinging a sword, and hammer. By exploring the applicability in more diverse VR scenarios, we will seek ways to maximize the VR experience. Such research is expected to greatly contribute to the development of VR technology.

### Acknowledgement

"This research was supported by the IITP under the Ministry of Science and ICT Korea through the ITRC program (IITP-2023-RS-2022-00156354).